# Bipolar electrocaloric effect in $Pb_{1-x}Gd_x(Mg_{1+x/3}Nb_{2-x/3})O_3$ ceramic: relaxor with super-dipolar glass state

Adityanarayan H. Pandey,[a,b,*] A.M. Awasthi,[c] and Surya Mohan Gupta[a,b,*]

[a]Homi Bhabha National Institute, Anushaktinagar, Mumbai-400094, India.
[b]Laser Materials Section, Raja Ramanna Centre for Advanced Technology, Indore-452013, India.
[c]UGC-DAE Consortium for Scientific Research, University Campus, Khandwa Road, Indore-452001, India.
*Email: anbp.phy@gmail.com; surya@rrcat.gov.in

The electrocaloric (EC) effect is calculated for $Pb_{1-x}Gd_x(Mg_{1+x/3}Nb_{2-x/3})O_3$; x = 0 to 0.1 relaxors from temperature dependent heat capacity and polarization measurements, using a thermodynamic Maxwell equation. Polarization change with temperature reveals an anomalous behaviour around the glass transition temperature for $x \geq 0.05$, which is in the vicinity of a crossover from positive to negative EC effect, which is explained evoking the critical slowing-down of the polar nano-domains dynamics to super-dipolar glass state upon cooling. The maximum negative EC coefficient $\xi_{max}$ is observed to decrease from ~0.4 K-mm/kV at 164 K to ~0.1 K-mm/kV with *x*.

Currently ferroelectric materials showing electrocaloric (EC) effect are in focus because of their promising usage for future generation cooling technologies as an alternative to the conventional vapour-compression refrigeration technology due to the absence of current-conduction in the ferroelectric materials [1]. Current impetus is on exploration of new ways to enhance the EC effect.

The EC effect ($\Delta T_{EC}$) has been first proposed in 1887 and experimentally observed 40 years later in the Rochelle Salt ($\Delta T_{EC}$ = 3 mK) [2]. Recent advancement in EC effect is due to the observation of giant EC response ($\Delta T_{EC}$ = 12K) around 222 °C in PZT ferroelectric thin films, and in ferroelectric/relaxor copolymer in temperature range 30-70 °C [3,4]. At present, several lead-based and non-lead based perovskite materials are known to show strong EC effect. Recently, an indirect measurement on lead magnesium niobate-lead titanate (PMN-PT) thin films has reported strong EC-response ($\Delta T_{EC}$ = ~31 K) for the applied voltage of 10 volt at 140 °C [5]. However, it is believed that only bulk materials with strong EC response would exhibit enough cooling power for new-generation cooling devices.

Ferroelectrics with large pyroelectric coefficient $(dP/dT)_E$ are generally known to exhibit strong EC response compared to the dielectrics. For a ferroelectric, temperature dependence of the EC effect ($\Delta T_{EC}$) at constant change of the electric field ($\Delta E$) shows a maximum at the transition temperature ($T_C$), where ferroelectric to paraelectric phase transition takes place. This positive EC effect is generally found in most of the ferroelectric materials, *i.e.*, FE-material heats up (cools down) with increase (removal) of the electric field [3-11]. Solid solution of the PMN-PT near the morphotropic phase boundary (MPB) is reported to show appreciable positive EC effect, due to the presence of polar nano-domains, which carry an extra entropy contribution pertinent to strong EC response [1].

Recently, negative EC effect is observed in relaxor ferroelectrics (RFE) and antiferroelectric (AFE) ceramics, where cooling is produced by the application of an electric field [12,13]. Similar negative EC effect and Existence of both positive and negative EC effects have been reported in a number of perovskites, solid state solutions, thin films, and multi-layered nanostructures [14-39]. Underlying mechanisms for the observance of the negative or both positive & negative EC effects is not yet established.

It has been previously reported that Gd-substitution in PMN ceramics enhances the dielectric relaxation strength, which has been correlated with enhancement in non-stoichiometric chemical ordered regions (CORs) [40-42]. Large concentration of Gd-ions at Pb-site is reported to induce critical slowing-down of dynamics of the polar nan-domains, which should cause anomalous EC effect, based on strong dependence of the dipolar entropy on the applied electric field [40-42]. The purpose of this study is to investigate the EC effect in $Pb_{1-x}Gd_x(Mg_{1+x/3}Nb_{2-x/3})O_3$, for $x = 0$, 0.01, 0.05 and 0.1 ceramics indirectly from the temperature dependent heat capacity and *P-E* hysteresis loop measurements on the ceramic compositions.

Ceramic samples of $Pb_{1-x}Gd_x(Mg_{1+x/3}Nb_{2-x/3})O_3$ (denoted as PGMN for $0 \leq x \leq 0.1$) have been synthesized similarly to the report [40-42]. Hewlett-Packart 4194A impedance analyzer is used to measure the dielectric response in the frequency range of 0.1 Hz - 100 kHz and temperature range of 120 - 450 K. Field induced polarization has been measured during heating at 50 Hz between 120 K and 350 K using Precision workstation of Radiant Technology, USA. Heat capacity ($C_p$) is measured using a commercial differential scanning calorimeter (STAR$^E$ DSC-1, Mettler Toledo) at the heating rate of 2 $^o$C/min.

A systematic study on Gd-doped PMN has already been reported, in which the Gd-ion is shown to substitute both at the Pb-site and the Mg-site for x ≥ 0.05 [40-42]. It has also been revealed that an additional disorder is induced when Gd-substitutes the Mg-site, resulting in higher value of degree of diffuseness $\delta_A$ (> 77) for x ≥ 0.05. This additional disorder is believed to be responsible for the reduction in the cooperative interaction among the PNRs, resulting in the critical slowing down of the PNRs dynamics, which ultimately leads to non-ergodic ferroelectric cluster glass ground state (also known as "super-dipolar" glass) [43]. Figure 1(a) depicts representative dielectric behaviour of x = 0.1 and inset of the Fig. 1(a) compares fitting of the $T_m$ vs $\tau$ plot to Eq. 1 for $0 \leq x \leq 0.1$.

$$\tau = \tau_o \left(\frac{T}{T_g} - 1\right)^{-zv} \quad (1)$$

where $\tau_o = \omega_o^{-1}$ is the microscopic time associated with flipping of fluctuating dipole, $zv$ is critical dynamic exponent and $T_g$ is glass transition temperature. As reported, reasonableness of the fitting parameters reveal interaction among the PNRs resulting into a critical slowing down of PNRs dynamics at finite temperature [41].

In order to evaluate the EC effect by "indirect method", heat capacity and *P-E* hysteresis loop is measured at different temperatures. Figure 1(b) compares temperature dependent heat capacity for $0 \leq x \leq 0.1$. The heat capacity initially varies linearly below ~ 160 K and then tends to saturate in the temperature range 160 K to 280 K. No structural phase transition is noticed for $0 \leq x \leq 0.1$. Figure 2 compares the *P-E* loops for *x* = 0 to 0.1 at several temperatures in 150 K to 240 K. Non-linear hysteresis-loss-free *P-E* loop is observed at 300 K for $0 \leq x \leq 0.1$. Small opening in the *P-E* loop at 210 K for $0.01 \leq x \leq 0.1$ is attributed to non-linear activity of sluggish PNRs response, in which the PNRs do not flip but just change shape, similar to a breathing mode reported for dipolar glass $K_{1-x}Li_xTaO_3$ [44]. For *x* < 0.05, the *P-E* loop at 300 K is "*s*-shape", which tends to saturate at the larger field and also exhibits non-linear P-E dependence. Below 210 K, a typical ferroelectric like *P-E* loop is observed for x = 0, indicating the development of long range order and the PMN is able to sustain $P_r$ at low temperatures. The coercive field increases with decreasing temperature, confirming the slowing down of PNRs' dynamics.

For $x \geq 0.05$, no transformation from short range (nano-domains) to long range (macroscale domains) ordering is noticed up to 50 kV/cm applied field. The maximum polarization ($P_{max}$) vs. applied field is observed to vary non-linearly for *x* < 0.05 and linearly for $x \geq 0.05$, signifying complete and incomplete alignment of the PNRs, respectively. Linear *P-E* dependence suggests that the correlation within the PNRs is weak, due to reduction in the size of the PNRs, rooted in the enhancement of the CORs and the presence of second phases e.g., MgO and $GdNbO_4$, as reported earlier [41].

Figure 3 shows the temperature dependence of the $P_{max}$ [$P_{max}$(T)] under different electric fields (5, 10, and 15 kV/cm), which is calculated using the upper/descending part of the hysteresis loops, similarly as reported by Mischenko *et. al.* [3]. The $P_{max}$(T) reveals a broad peak, which shifts linearly toward higher temperature with increasing electric field for $x \geq 0.01$. This anomalous behaviour is observed in very few materials, which show re-entrant behaviour [45-48]. Field-induced alignment of the PNRs will depend upon the combination of applied field and thermal agitation or dynamics of PNRs. Higher field (15 kV/cm) should overcome the agitation effect and cause alignment of the PNRs at higher temperature, compared to that at lower field (5 kV/cm). It is believed that the anomalous behaviour observed for $0.01 \leq x \leq 0.1$ associates with the dynamics of polar nano-domains. Two distinct regions (left and right of the peak) are noticed in Figs. 3(a-d). In region II (right of the peak, high temperature side), the $P_{max}$ is observed to increase with cooling, which is attributed to the enhanced alignment of the PNRs in the direction of electric field. Depending upon the composition, long range ordering is developed for $x = 0$, but the extent of the co-operative interaction decreases with increase in the Gd-ions concentration. For $x \geq 0.5$, no long range ordering is observed, as shown in Fig. 2(c-d); an increase in the $P_{max}$ value with decreasing temperature is attributed to the alignment of smaller sized nano-domains. The $P_{max}$ decreases with further cooling in region I, which is attributed to the non-ergodic behaviour and critical slowing down of the PNR's dynamics for $x = 0$ and $x \geq 0.05$, respectively. The critical slowing-down of the PNR's dynamics is due to the enhanced disorder observed with the Gd-ion substitution at the Mg-site for $x \geq 0.5$. It seems that the Gd-ion substitution at the Mg-site enhances sluggish alignment of the PNRs in the direction of the applied electric field below a glass transition temperature, $T_g$.

The EC effect is adiabatic temperature change ($\Delta T_{EC}$) of a dielectric material in response to the change in electric field ($\Delta E$). Assuming the Maxwell's relation $(\partial P/\partial T)_E = (\partial S/\partial E)_T$, for a dielectric material having density ($\rho$) and heat capacity ($C_p$), $\Delta T_{EC}$ due to an applied field $E$ is given as follows [1]

$$\Delta T_{EC} = -\frac{1}{\rho} \int_{E_1}^{E_2} \frac{T}{C_p(T)} \left(\frac{\partial P}{\partial T}\right)_E dE \qquad (2)$$

where $P$ is the polarization, and $E_1$, $E_2$ are respectively the starting and final applied electric fields. The values of $(\partial P/\partial T)_E$ are calculated from the derivative of $P$ vs. $T$ plot (Fig. 3). For the density '$\rho$' (= mass of unit cell/(lattice constant)$^3$) of the ceramic, lattice constant is determined from the Rietveld refinement of XRD pattern using Fullprof software ($\rho \sim 8.066$ gm/cm$^3$) [41].

The $\Delta T_{EC}$ is calculated by using Eq. 2, where $\Delta E$ (=$E_2-E_1$) is 5, 10, and 15 kV/cm when $E_1$ is set as zero. Figure 4(a-d) shows $\Delta T_{EC}$ as a function of temperature at 5, 10, and 15 kV/cm values of $\Delta E$ for $0 \leq x \leq 0.1$. Earlier, Rozic *et. al.* [49,50] reported the EC effect of PMN in the temperature range of 220-340 K and found consistent $\Delta T_{EC}$ value with that of $x$ =0. But as a consequence of the anomalous polarization-change with temperature, all PGMN ceramics have exhibited a crossover from positive to negative $\Delta T_{EC}$ in the vicinity of the temperature dependent $P_{max}$-peak, as observed in Figs. 3(a-d). Figure 4(e) compares temperature dependent $\Delta T_{EC}$ at 15 kV/cm for $0 \leq x \leq 0.1$ showing shifting of maximum/minimum positive/negative values of $\Delta T_{EC}$ with increasing "x".

The negative EC effect in region I is the outcome of decrease of polarization with cooling where $(\partial P/\partial T)_E$ decreases. According to the Maxwell relation $(\partial P/\partial T)_E = (\partial S/\partial E)_T$, the entropy should decrease with the applied field upon cooling, hence the decrease of temperature is observed. Moreover, the crossover temperature ($T_{co}$) of $\Delta T_{EC}$ shifts toward higher temperature with increase in electric field. A shift of large $\Delta T_{co} \sim 24$ K is observed as the electric field is increased from 5 to 15 kV/cm. In contrast to this, in the case of AFE La-doped PZT thin films, Geng *et. al.* [20] reported that the $T_{co}$ shifts to lower temperature side with increase of electric field. However, Bhaumik *et. al.* [22] and Ramesh *et. al.* [30] have observed similar analogous behaviour of $T_{co}$ with the field. Also, Zhou *et. al.* [23,24] reported the coexistence of multiple negative and positive EC effects in (Pb,La)(Zr,Sn,Ti)O$_3$ single crystals. The anomalous crossover and negative EC effect is recently reported in many AFEs and RFEs [12-39]. However, the mechanisms for the negative EC effect is not yet clearly understood.

A negative EC effect is expected if there is a transition between the two states of the system, where the lower-temperature state has a smaller polarisation than the higher-temperature state. Generally, electric-field-induced phase transition is believed to relate with the negative $\Delta T_{EC}$. The negative EC effect is also reported in PMN-PT crystal and attributed to field-induced structural transformation from monoclinic to orthorhombic phase [14]. In the case of Na$_{0.5}$Bi$_{0.45}$TiO$_3$ (NBT), formation of incommensurate AFE phase is responsible for the negative EC effect, where application of electric field favours AFE or increased dipolar disorder [12,13]. Statistical mechanics based microscopic model by Axelsson *et. al.* [35,36] implied two phase transitions, which are very close in temperature, resulting in the coexistence of a dual-nature EC effect in PMN-PT. The first-principles based simulations by Ponomareva *et. al.* [38] attributed the non-linearity between the P-E as the origin of negative EC effect in $Ba_{0.5}Sr_{0.5}TiO_3$. Moreover, there is a broad minimum value of $\Delta T_{EC}$ ~ -0.15 K, observed under an applied field of 15 kV/cm towards negative region around 152 K. A remarkable feature is that the position of negative minimum temperature is independent of the applied electric field. Recently, Wu *et. al.* [33] demonstrated that high efficiency in EC based solid state cooling devices is achieved through coexistence of positive and negative EC effects near pseudo first order phase transition in ferroelectrics. For practical applications, EC materials show that change of small electric field generates large temperature change. The strength of EC effect, i.e., responsivity ($\xi_{max}$), is calculated according to $\xi_{max} = \Delta T_{ECmax}/\Delta E_{max}$ where $\Delta T_{ECmax}$ is the maximum temperature change and $\Delta E_{max}$ is the corresponding electric field change. The value of maximum negative EC coefficient ($\xi_{max}$) is observed to decrease from ~ 0.4 K-mm/kV at 164 K to ~ 0.1 K-mm/kV with increasing Gd-content. The $\xi_{max}$ ~ 0.2 K-mm/kV at 150 K for PMN is consistent with earlier report [49,50]. In contrast, $\xi_{max}$ decreases consistently in region I with increase of Gd-substitution. Similar EC behaviour i.e., crossover from positive to negative values near the Vogel-Fulcher freezing temperature is reported for $Pb_{0.8}Ba_{0.2}[(Zn_{1/3}Nb_{2/3})_{0.7}Ti_{0.3}]O_3$ and Gd-doped $Na_{0.5}Bi_{0.45}TiO_3$ [27,39].

It has already been reported [41] that 5 at.% Gd-doped PMN has both the PNRs and the CORs, which are randomly distributed within the grain, which tend to align along the external electric field. Above room temperature, thermal energy overcomes alignment of the PNRs and remnant polarization are not sustained. However, at low temperatures, thermal energy is not sufficient to disrupt this alignment, but below a certain temperature ($T_g$), the dynamics of the PNRs critically slow down causing incomplete alignment, which results in lower polarization and crossover from positive to negative $\Delta T_{EC}$. Further decrease in temperature results in more out-of-phase response of the PNRs, leading to a minimal near 150 K. Therefore, the crossover between positive to negative EC is believed to relate to the critical slowing down of PNRs into super-dipolar glass state. At present, it is not clear if the CORs are responding to the external electric field and how this response varies with temperature. Increase in $\Delta T_{EC}$ below 150 K suggests active role of the CORs, which requires further confirmation.

In conclusion, the EC effect exhibited an anomalous crossover from low temperature negative EC to high temperature positive EC effect near the $T_g$ ~204 K at 5 kV/cm and shifting of crossover temperature with increasing electric field for $x \geq 0.05$ is explained by critical slow-down glassy dynamics of polar nano-domains. The presence of bipolar EC effect makes it a promising material for refrigeration technologies. The maximum value of negative EC coefficient (ξ) for "x" = 0.01 is 0.4 K-mm/kV at 195 K.

A.H. Pandey acknowledges Suresh Bahardwaj for heat capacity measurement. Homi Bhabha National Institute, India is acknowledged for research fellowship of AHP.

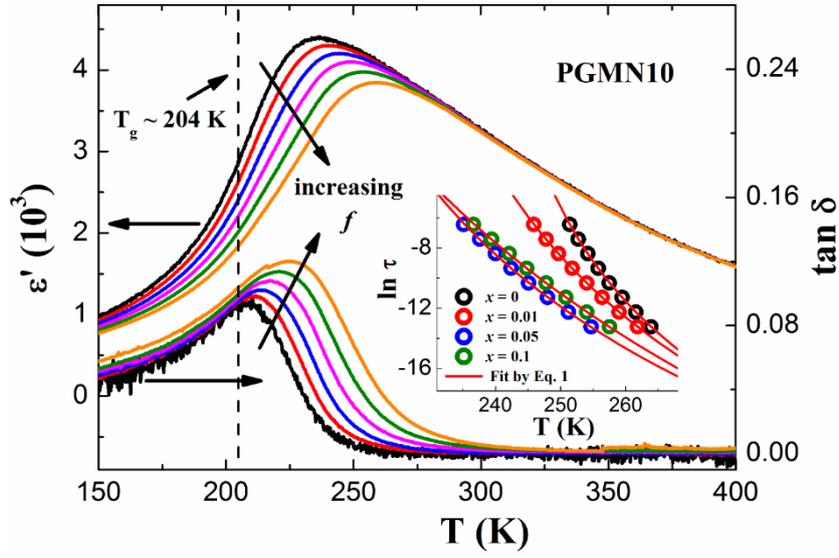

(a)

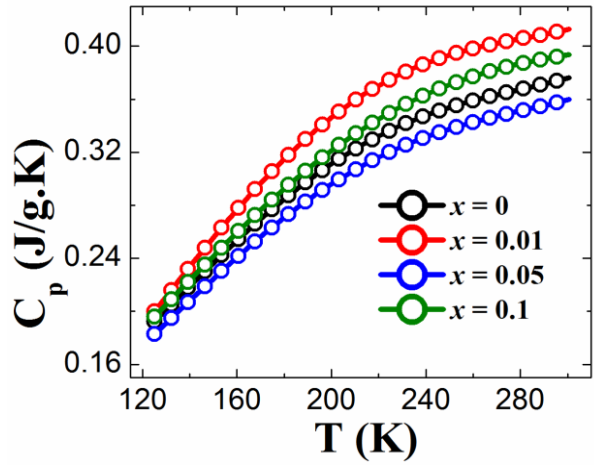

(b)

**Figure 1.** Temperature dependent (a) *ε' and tanδ* and of $Pb_{1-x}Gd_xMg_{1+x/3}Nb_{2-x/3}O_3$ ceramic of *x*=0.1 at different frequency (b) Heat capacity ($C_p$) for $Pb_{1-x}Gd_xMg_{1+x/3}Nb_{2-x/3}O_3$ ($0 \leq x \leq 0.1$). Inset of the Fig. 1(a) shows fitting of all PGMN ceramics to the Eq. 1.

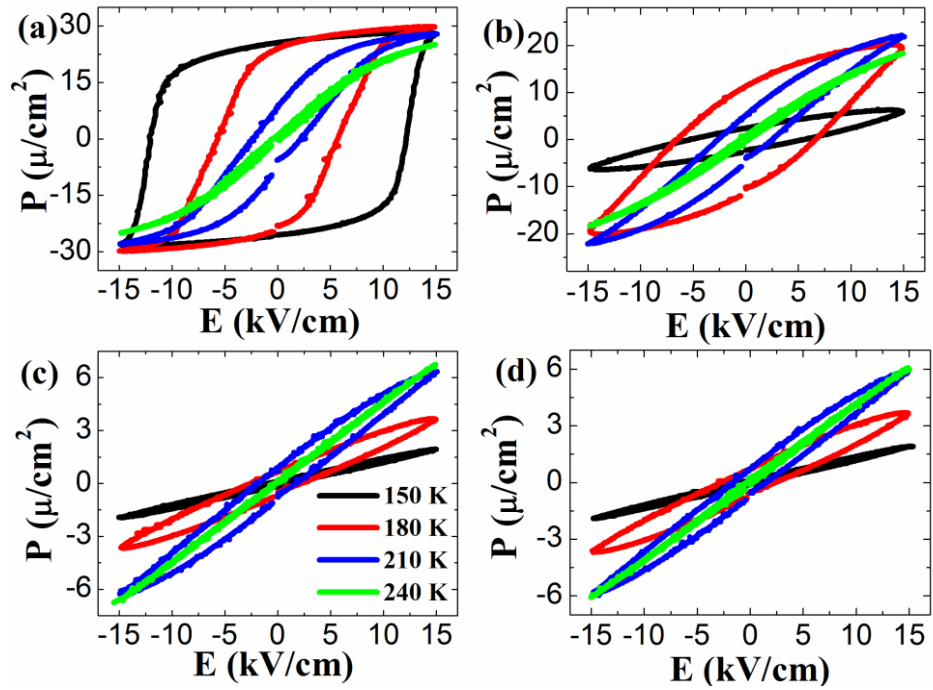

**Figure 2.** Temperature dependent *P-E* hysteresis loop of Pb$_{1-x}$Gd$_x$Mg$_{1+x/3}$Nb$_{2-x/3}$O$_3$ ceramics recorded at 50 Hz; **(a)** *x* =0, **(b)** *x* = 0.01, **(c)** *x* = 0.05, and **(d)** *x* = 0.1.

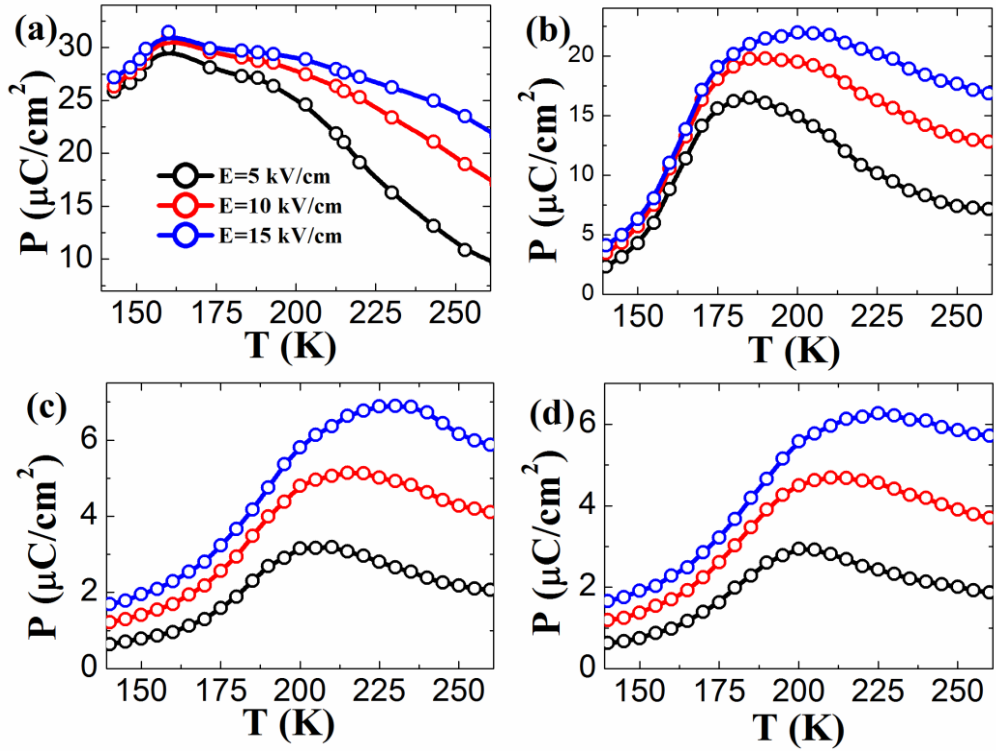

**Figure 3.** Temperature dependent polarization behaviour of $Pb_{1-x}Gd_xMg_{1+x/3}Nb_{2-x/3}O_3$ ceramics at different electric fields; **(a)** $x = 0$, **(b)** $x = 0.01$, **(c)** $x = 0.05$, and **(d)** $x = 0.1$.

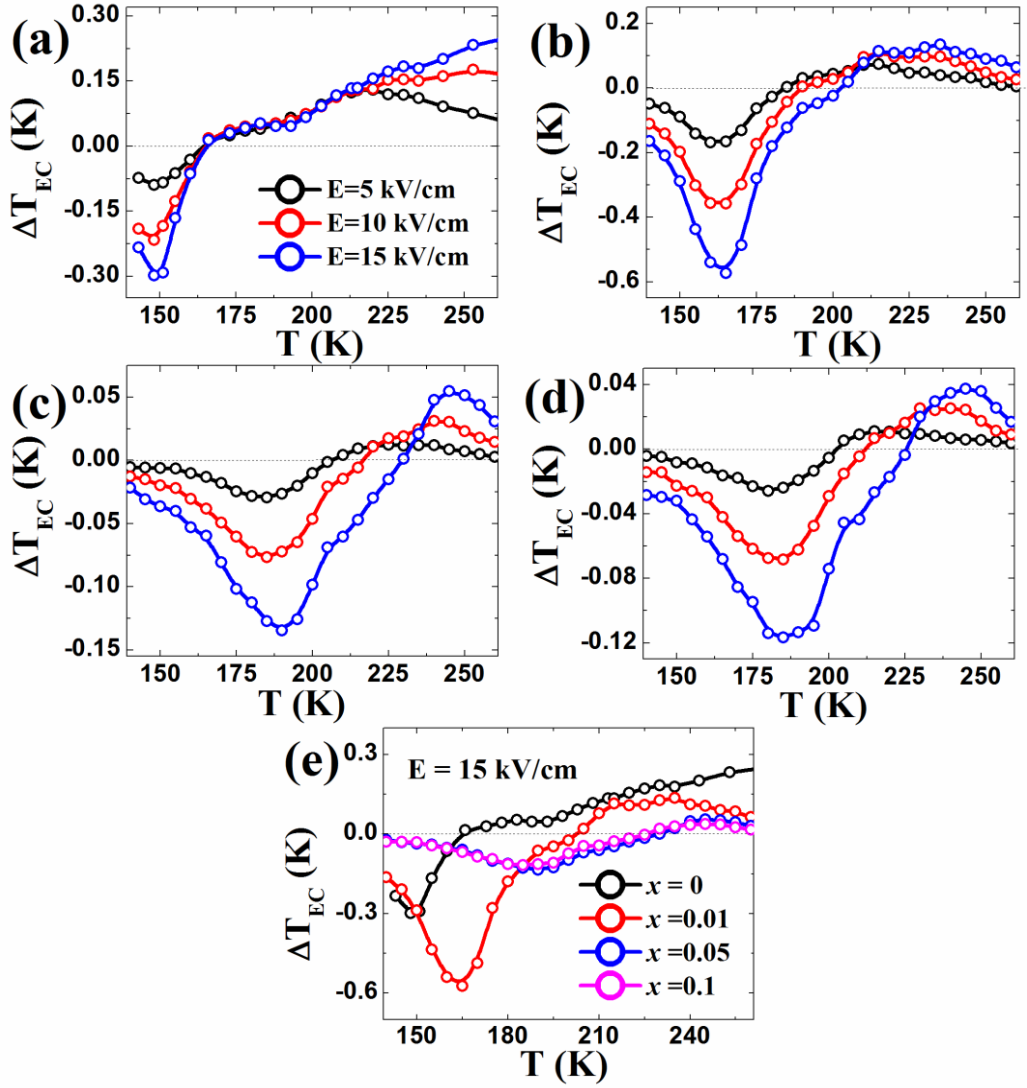

**Figure 4.** Electrocaloric temperature change ($\Delta T_{EC}$) as a function of temperature at different applied electric field for $Pb_{1-x}Gd_xMg_{1+x/3}Nb_{2-x/3}O_3$ ceramics **(a)** $x = 0$, **(b)** $x = 0.01$, **(c)** $x = 0.05$, and **(d)** $x = 0.1$, **(e)** comparison of $\Delta T_{EC}$ as a function of temperature for different compositions ($0 \leq x \leq 0.1$) at $E = 15$ kV/cm